\documentclass{WileyMSP-template}

\usepackage[utf8]{inputenc}
\usepackage[T1]{fontenc}

\usepackage[superscript,nomove]{cite}
\makeatletter\renewcommand\@citess[1]{\textsuperscript{[#1]}}\makeatother

\usepackage{mathtools}
\usepackage{amssymb}
\usepackage{bm}
\usepackage{siunitx}
\usepackage{booktabs}
\usepackage{csquotes}

\DeclareMathOperator{\arsinh}{arsinh}\renewcommand*{\arsinh}{\sinh^{-1}}
\DeclareMathOperator{\arcosh}{arcosh}\renewcommand*{\arcosh}{\cosh^{-1}}
\DeclareMathOperator{\artanh}{artanh}\renewcommand*{\artanh}{\tanh^{-1}}

\DeclarePairedDelimiter{\parens}{\lparen}{\rparen}
\DeclarePairedDelimiter{\bracks}{\lbrack}{\rbrack}

\let\abs\verts

\newcommand*{\x}{\mathclose{}\mathopen{\mskip\thinmuskip}}
\newcommand*{\diff}[1]{\mathrm{d}#1}
\newcommand*{\derivop}[1]{\partial_{#1}}
\newcommand*{\deriv}[2]{\frac{\mathopen{\partial}#1}{\mathopen{\partial}#2}}

\newcommand*{\transpose}[1]{#1^{\mathsf{T}}}
\NewDocumentCommand{\tup}{smo}{\IfNoValueTF{#3}{\bm{#2}}{\IfBooleanTF{#1}{{#2}_{#3}}{{#2}^{#3}}}}
\DeclarePairedDelimiterX{\tupdelims}[1]{\lparen}{\rparen}{\cramped{#1}}
\makeatletter\newcommand*{\tupderivop}[1]{\tup{\nabla}_{\mkern-\medmuskip #1}}\makeatother

\DeclarePairedDelimiterX{\setdelims}[1]{\lbrace}{\rbrace}{\cramped{#1}}

\DeclarePairedDelimiterXPP{\func}[2]{#1}{\lparen}{\rparen}{}{\cramped{#2}}

\newcommand*{\const}{\mathrm{const}}
\newcommand*{\incr}[1]{\Delta #1}
\newcommand*{\hgf}{{}_2 F_1}

\NewDocumentCommand{\drift}{om}{\tup{\mu}[#1]_{#2}}
\NewDocumentCommand{\diffusion}{om}{\tup{\sigma}[#1]_{#2}}
\NewDocumentCommand{\diffusivity}{om}{\tup{D}[#1]_{#2}}
\newcommand*{\pdf}[1]{f_{#1}}
\newcommand*{\fpo}[1]{L_{#1}}
\newcommand*{\potential}[1]{\Phi_{#1}}

\newcommand*{\ini}{\mathrm{ini}}
\newcommand*{\fin}{\mathrm{fin}}
\newcommand*{\equ}{\mathrm{equ}}

\newcommand*{\trv}{\mathrm{T}}
\newcommand*{\lgt}{\mathrm{L}}
\newcommand*{\mT}{m_{\trv}}
\newcommand*{\pT}{p_{\trv}}
\newcommand*{\PT}{P_{\trv}}
\newcommand*{\pL}{p_{\lgt}}
\newcommand*{\PL}{P_{\lgt}}
\NewDocumentCommand{\rap}{o}{\tup{\xi}[#1]}
\NewDocumentCommand{\Rap}{o}{\tup{\Xi}[#1]}
\newcommand*{\DT}{D_{\trv}}
\newcommand*{\DL}{D_{\lgt}}

\newcommand*{\ysurf}{y_{\mathrm{s}}}
\newcommand*{\betasurf}{\beta_{\mathrm{s}}}

\newcommand*{\sqrtsNN}{\sqrt{s_{\mathrm{NN}}}}
\newcommand*{\yb}{y_{\mathrm{b}}}
\newcommand*{\mN}{m_{\mathrm{N}}}
\newcommand*{\Qs}{Q_{\mathrm{s}}}

\newcommand*{\gluglu}{\mathrm{gg}}
\newcommand*{\quaglu}{\mathrm{qg}}
\newcommand*{\gluqua}{\mathrm{gq}}

\newcommand*{\cpion}{\mathrm{\pi}^\pm}
\newcommand*{\ckaon}{\mathrm{K}^\pm}
\newcommand*{\cnucleon}{\mathrm{N}^\pm}

\newcommand*{\dNdeta}{\diff{N} / \diff{\eta}}
\newcommand*{\dNdpTdeta}{\diff{^2 N} / \parens{\diff{\pT} \x \diff{\eta}}}

\usepackage{hyperref}
\usepackage[capitalize,noabbrev]{cleveref}

\begin{document}

\pagestyle{fancy}
\rhead{\includegraphics[width=2.5cm]{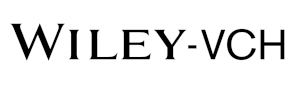}}

\title{Cylindrically symmetric diffusion model\\for relativistic heavy-ion collisions}
\maketitle

\author{Johannes Hoelck}
\author{Georg Wolschin*}


\begin{affiliations}
Institute for Theoretical Physics, Heidelberg University\\
Philosophenweg~16\\
69120~Heidelberg\\
Baden-Württemberg, Germany\\
*Email Address:
wolschin@thphys.uni-heidelberg.de
\end{affiliations}


\begin{abstract}
A relativistic diffusion model with cylindrical symmetry, which propagates an initial state based on quantum chromodynamics in time towards a thermal equilibrium limit, is derived from nonequilibrium--statistical considerations:
Adapting an existing framework for Markovian stochastic processes representing relativistic phase-space trajectories, a Fokker--Planck equation is obtained for the time evolution of particle-number distribution functions with respect to transverse and longitudinal rapidity.
The resulting partially-evolved distribution functions are transformed to transverse-momentum and pseudorapidity space, and compared with charged-hadron data from the CERN Large Hadron Collider (LHC).
\end{abstract}



\section{Introduction}
\label{sec:introduction}

In most of the available models and theories of relativistic heavy-ion collisions, charged-hadron production is considered to occur from the hot fireball of partons that expands, cools, and eventually hadronizes in a parton--hadron crossover.
Typically, a well-defined initial state such as the color-glass condensate (CGC) is matched smoothly to viscous hydrodynamics \cite{GaleEtAl2013PRL110}, to finally use Cooper--Frye freezeout or another code that accounts for the final-state interactions.
In a nonequilibrium--statistical relativistic diffusion model (RDM) with three sources \cite{Wolschin2015PRC91}, the highly excited fragments also partake in particle production, in addition to the central fireball source.

These earlier phenomenological diffusion-model calculations for relativistic heavy-ion collisions were mostly done in rapidity (\(y\)) space for stopping (baryon minus antibaryon) distributions, or --~following a Jacobian transformation~-- in pseudorapidity (\(\eta\)) space for produced charged hadrons.
The results were compared to the corresponding experimental marginal spectra if available, or with narrow cutouts from joint spectra with additional transverse-momentum dependence (\(\dNdpTdeta\)), thereby determining the free parameters of the model, such as the diffusivity, the equilibration times in the individual sources, and the particle numbers in least-squares fits.

The purpose of the present work is twofold:
We place the relativistic diffusion model on a firm nonequilibrium--statistical foundation by using the mathematics of stochastic calculus for its formulation, and design the model using both transverse and longitudinal degrees of freedom.
It is eventually formulated in three dimensions with cylindrical symmetry, implemented by a two-dimensional parameterization of momentum space, and compared to recent LHC data.
The increase in dimensionality offers the advantage that both marginal and joint particle distributions can now be computed natively within the model.
We maintain the distinction between three sources of particle production, while also incorporating the varying masses of different particle species.

While the measurements of negatively plus positively charged hadrons cannot distinguish particles that are produced from the central fireball versus the ones arising from the two fragmentation sources, the latter are clearly visible in net-baryon (\enquote{stopping}) distributions that can be deduced from proton minus antiproton data.
Here, the central source cancels out because baryons and antibaryons are produced in equal amounts, and the fragmentation peaks become measurable in net-proton data.
The stopping measurements require, however, to distinguish protons and antiprotons from pions at very forward angles -- which is so far possible only at energies reached at the Super Proton Synchrotron (SPS), and to some extent at the Relativistic Heavy Ion Collider (RHIC), but not yet at the Large Hadron Collider (LHC).

In earlier works \cite{Mehtar-TaniWolschin2009PRL102,Mehtar-TaniWolschin2009PRC80}, we had accounted for the role of the fragmentation sources that are visible in stopping using a model that relies on gluon-saturation physics, where the valence quarks in the forward-going ion interact with soft gluons in the backward-going collision partner, thus slowing down the fragments.
The positions of the stopping peaks in rapidity space were shown to depend upon the gluon saturation momentum, offering the possibility to measure it through a determination of the gluon saturation-scale exponent once net-proton forward-angle data would become available in heavy-ion collisions at LHC energies.

These calculations were done in the color-glass condensate framework based on small-coupling quantum chromodynamics (QCD).
We have incorporated this approach into a nonequilibrium--statistical model \cite{HoelckWolschin2020PRR2} that also accounts for the time-dependence of the stopping process.
The combination of the CGC-based model with a relativistic diffusion approach becomes possible through a suitably designed fluctuation--dissipation relation (FDR).
This allows to devise a fully time-dependent theory that draws aspects from QCD.
An existing general framework for relativistic stochastic drift--diffusion processes in phase space \cite{DebbaschEtAl1997JSP88,DunkelHaenggi2005PRE71,DunkelHaenggi2005PRE72,DunkelHaenggi2009PR471} has been used to derive a Fokker--Planck equation (FPE) in rapidity space, which provided a firm mathematical basis for the earlier phenomenological models and data analyses \cite{Wolschin1999EPJA5,Wolschin2015PRC91,Wolschin2016PRC94}.

Within this approach, the initial state in each nucleus was prepared as a zero-temperature Fermi gas, from which the initial probability density in rapidity space can be calculated analytically.
The transport coefficients of the FPE that governs the time evolution of the density have been determined such that the solution approaches the CGC distribution \cite{Mehtar-TaniWolschin2009PRL102}.
This is achieved by linking the drift and diffusion coefficients by a fluctuation--dissipation relation obtained from the logarithmic derivative of the color-glass probability density function \cite{HoelckWolschin2020PRR2}.
The resulting equation can be solved numerically and yields remarkable agreement with the available stopping data from SPS and RHIC.

In a subsequent publication \cite{HoelckEtAl2023PLB840}, we have used this mesoscopic model to show that the net-proton rapidity spectra at very forward angles exhibit a scaling behavior reminiscent of --~but different from~-- limiting fragmentation that is related to geometric scaling in the color-glass condensate, and depends upon the gluon saturation scale.
This scaling is in agreement with SPS and RHIC net-baryon data from \(\sqrtsNN = \qtyrange{6.3}{200}{\GeV}\), but a comparison at LHC energies requires the future availability of such data.

The success of the relativistic diffusion model to account for the stopping of relativistic heavy ions motivates its extension to thermalization in particle production.
Here, the central fireball becomes the main source of charged-hadron production, mostly through gluon--gluon collisions -- in addition to the particles that are generated from the two fragmentation sources.
While early calculations in a three-source model based on a FPE in rapidity space with phenomenological values for the drift and diffusion coefficients already resulted in agreement with centrality-dependent pseudorapidity data from RHIC and LHC --~and also offered reliable predictions at higher energies \cite{BiyajimaEtAl2002PTP108,WolschinEtAl2006AP518,WolschinEtAl2006PLB633,Wolschin2013JPG40,Wolschin2015PRC91,Wolschin2016PRC94}~--, we now present a diffusion model for particle production based on a rigorous derivation of the underlying transport equation from a consistent relativistic theory, and extend it to simultaneously treat the transverse and longitudinal degrees of freedom in momentum space.

Here, the form of the initial state is determined from QCD, analogous to the stationary state in stopping.
For the asymptotic equilibrium state, it is expected that the majority of the produced particles participates in the thermalization process, and we choose thermal distributions for all three sources to model the equilibrium state, and thereby the fluctuation--dissipation relations for the produced hadrons.
Since the system is not spatially confined, it cannot reach a true thermal equilibrium that is stationary in position space and independent of its initial configuration.
Therefore, we use modified thermal distributions that exhibit a collective expansion of the system.
Moreover, the high-momentum tails are adapted to take the observed transition to power-law momentum distributions properly into account.
With the transport coefficients derived mesoscopically within this framework, we solve the FPE numerically to obtain marginal rapidity spectra and joint transverse-momentum and pseudorapidity spectra for produced charged particles such as pions, kaons, and protons, which can directly be compared to available central Pb--Pb collision data from the ALICE \cite{ALICE2013PLB726,ALICE2017PLB772,ALICE2018JHEP2018} and ATLAS \cite{ATLAS2015JHEP2015} collaborations.

The relativistic diffusion model for particle production in momentum space is derived from the underlying stochastic relativistic theory in \cref{sec:model}.
Using a Kramers--Moyal expansion, a Fokker--Planck equation in two-dimensional rapidity space is obtained.
Initial and equilibrium states are accounted for in detail, and the numerical method to solve the FPE in two dimensions is outlined.
Initial states for the three sources are calculated from the distribution functions of the partons characterizing the respective particle production process.
The equilibrium distribution that is reached asymptotically for \(t \to \infty\) is taken to be a thermal distribution including collective expansion.

The results are presented in \cref{sec:results} for marginal pseudorapidity spectra of produced pions, kaons, protons, and their antiparticles, as well as for the corresponding joint transverse-momentum and pseudorapidity spectra in central Pb--Pb collisions.
Direct comparisons with ALICE charged-hadron production data at \qtylist{2.76;5.02}{\TeV} are presented.
The role of fragmentation versus central sources in particle production is discussed.
Calculated joint distribution functions are directly compared to recent ATLAS measurements, and the normalized residuals are determined in two dimensions.

The conclusions and perspectives of this investigation are presented in \cref{sec:conclusion}.

\section{Model}
\label{sec:model}

Capturing the numerous particle interactions during a heavy-ion collision in the relatively simple mathematical form of a drift--diffusion process is a difficult
--~if not impossible~--
task.
The scope of our current work therefore does not consist in providing a fully realistic picture of the microscopic dynamics, but in \enquote{finding the \enquote{best} approximation of the exact dynamics}\cite{DunkelHaenggi2009PR471} within the phase-space diffusion formalism.
This model is then used to draw conclusions about the observed macroscopic behavior of the system in direct comparisons with experimental data, which we take from available recent results of the ALICE and ATLAS collaborations at the Large Hadron Collider.

We will make heavy use of the parton--hadron duality throughout this article by assuming a continuity between the trajectories of partonic pseudoparticles in the fireball of deconfined matter created shortly after the initial binary nucleon collisions and those of proper hadronic particles in the ensuing hadron-gas phase.
The masses of pseudoparticle
--~understood as a group of loosely connected partons~--
and corresponding particle are assumed to be identical, and both are denoted by the same symbol~\(m\).

\subsection{Stochastic drift--diffusion process}

We model the charged-particle trajectories as stochastic drift--diffusion processes that follow from generalization of the concept of relativistic Brownian motion \cite{DebbaschEtAl1997JSP88,DunkelHaenggi2005PRE71,DunkelHaenggi2005PRE72}.
The latter are formulated with respect to the Brownian particle's phase-space configuration~\(\tupdelims{\tup{x},\tup{p}}\) instead of merely its position~\(\tup{x}\), which is a consequence of the fundamental mathematical principle that stochastic processes in position space cannot be both Lorentz-invariant and (first-order) Markovian \cite{Lopuszanski1953APP12,Dudley1966AM6,Hakim1968JMP9}.
By including the particle's momenta~\(\tup{p}\) in the definition of the trajectory, the resulting drift--diffusion process is second-order Markovian when expressed purely in position-space coordinates, which allows to circumvent the aforementioned problem while keeping the beneficial mathematical properties of Markov processes.

In the following, \(\tup{X} = \tupdelims{\tup{X}[1],\tup{X}[2],\tup{X}[3]}\) and \(\tup{P} = \tupdelims{\tup{P}[1],\tup{P}[2],\tup{P}[3]}\) denote the probabilistic trajectories of the Brownian particle in position and momentum space, respectively.
Increments~\(\diff{\func{\tup{X}[\mu]}{t}} \coloneq \func{\tup{X}[\mu]}{t + \diff{t}} - \func{\tup{X}[\mu]}{t}\) of the position-space trajectory due to an infinitesimal timestep~\(\diff{t}\) in the laboratory frame obey the stochastic differential equation (SDE) \cite{discretization_scheme}
\begin{equation}
	\label{eq:SDE-position}
	\diff{\func{\tup{X}[\mu]}{t}} = \frac{\func{\tup{P}[\mu]}{t}}{\func{\tup{P}[0]}{t}} \x \diff{t}
	\quad\text{for}\quad
	\mu = 0,\dots,3
\end{equation}
with the particle's energy \cite{natural_units}
\begin{equation}
	\label{eq:SDE-mass-shell}
	\func{\tup{P}[0]}{t} \coloneq \sqrt{\textstyle m^2 + \sum_{i=1}^3 \func{\tup{P}[i]}{t}^2}
	\mathinner{,}
\end{equation}
which is fixed by the mass-shell condition.
With respect to the coordinate axes, we follow the common convention to place the \(3\)-axis parallel to the beam axis so that \(\tup{x}[1]\) and \(\tup{x}[2]\) span the so-called transverse plane.

Regarding the time-evolution of the particle's momentum, we assume \(\tup{P}\) to follow a drift--diffusion SDE,
\begin{equation}
	\label{eq:SDE-momentum}
	\diff{\func{\tup{P}[i]}{t}} =
	\func[\big]{\drift[i]{\tup{P}}}{\func{\tup{P}}{t}} \x \diff{t}
	+
	\sum_{k=1}^{3} \func[\big]{\diffusion[i k]{\tup{P}}}{\func{\tup{P}}{t}} \x \diff{\func{\tup{W}[k]}{t}}
	\quad\text{for}\quad
	i = 1,2,3
	\mathinner{,}
\end{equation}
which is characterized by a \(3\)-dimensional drift coefficient~\(\drift{\tup{P}}\) and a (\(3 \times 3\))-dimensional diffusion coefficient~\(\diffusion{\tup{P}}\) that represent generalized directed (deterministic) and undirected (stochastic) forces, respectively.
The latter are driven by the increments of a \(3\)-dimensional standard Wiener process~\(\func{\tup{W}}{t}\), the mathematical formulation of Gaussian white noise.

In \cref{eq:SDE-momentum}, we assumed drift and diffusion
to be independent of the particle's current position~\(\func{\tup{X}}{t}\) \cite{position_independence}, which is equivalent to neglecting any external forces.
The entire dynamics of the system is then encoded in the momentum-space process~\(\tup{P}\) so that the position-space process~\(\tup{X}\) can be disregarded in the following discussion.

Further, we only consider central heavy-ion collisions with small impact parameter in this article where the system has a rotational symmetry with respect to the beam axis in good approximation, so that our problem becomes effectively two-dimensional when expressed in terms of transverse momentum~\(\pT \coloneq \sqrt{\parens{\tup{p}[1]}^2 + \parens{\tup{p}[2]}^2}\) and longitudinal momentum~\(\pL \coloneq \tup{p}[3]\).
The transverse-plane angular coordinate can then be eliminated from \cref{eq:SDE-momentum}, which results in the structurally similar drift--diffusion SDE
\begin{equation}
	\label{eq:SDE-momentum-TL}
	\diff{\func{\tup{\hat{P}}[i]}{t}} =
	\func[\big]{\drift[i]{\tup{\hat{P}}}}{\func{\tup{\hat{P}}}{t}} \x \diff{t}
	+
	\sum_{k=1}^{2} \func[\big]{\diffusion[i k]{\tup{\hat{P}}}}{\func{\tup{\hat{P}}}{t}} \x \diff{\func{\tup{W}[k]}{t}}
	\quad\text{for}\quad
	i = 1,2
\end{equation}
with \(\tup{\hat{P}} \coloneq \tupdelims{\PT,\PL}\) and associated \(2\)-dimensional drift and (\(2 \times 2\))-dimensional diffusion coefficient functions \(\drift{\tup{\hat{P}}}\) and \(\diffusion{\tup{\hat{P}}}\), respectively.
The latter can be connected to \(\drift{\tup{P}}\) and \(\diffusion{\tup{P}}\) via stochastic calculus, as detailed in \cref{app:coefficient-transformation}.

To simplify Lorentz boosts along the beam axis and to compress the support of the PDF \cite{hyperbolic_transformation}, it is convenient to perform an additional, hyperbolic transformation, that yields the coordinates
\begin{equation}
	\label{eq:rapidity-coordinates}
	h \coloneq \func[\big]{\arsinh}{\pT / m}
	\qquad \text{and} \qquad
	y \coloneq \func[\big]{\artanh}{\pL / E}
	\mathinner{,}
\end{equation}
which we will denote as transverse and longitudinal rapidity, respectively, in accordance with their connection to the transverse and longitudinal momentum.
The transverse rapidity~\(h\) maps low \(\pT\) to a linear and high \(\pT\) to a logarithmic scale; it is mathematically equivalent to a coordinate of the same name introduced earlier for the analysis of scaling behavior in experimental data \cite{TrainorPrindle2016PRD93}.
Conveniently, the expression \(m \x \func{\cosh}{h} = \sqrt{m^2 + \pT^2} \eqcolon \mT\) gives the particle's transverse mass.
In these new coordinates, the drift--diffusion SDE~\labelcref{eq:SDE-momentum-TL} takes the form
\begin{equation}
	\label{eq:SDE-rapidity}
	\diff{\func{\Rap[i]}{t}} =
	\func[\big]{\drift[i]{\Rap}}{\func{\Rap}{t}} \x \diff{t}
	+
	\sum_{k=1}^{2} \func[\big]{\diffusion[i k]{\Rap}}{\func{\Rap}{t}} \x \diff{\func{\tup{W}[k]}{t}}
	\quad\text{for}\quad
	i = 1,2
	\mathinner{,}
\end{equation}
where we use the symbol~\(\Rap \coloneq \tupdelims{H,Y}\) to denote the transformed \(2\)-dimensional drift--diffusion process in transverse and longitudinal rapidity space.

To determine the dynamics of the Brownian particle, it is now sufficient to specify the functional form of the drift and diffusion coefficients for one of the SDEs presented above.
This automatically fixes the form of the other SDEs, as they are all interconnected via stochastic calculus.

For the time being, we choose the rapidity-space diffusion coefficient~\(\diffusion{\Rap}\) to be constant and the matrix product~\(\diffusion{\Rap} \x \transpose{\diffusion{\Rap}}\) to be diagonal, in order to simplify the numerical treatment.
While the nonconstant, nondiagonal case is accessible within our model framework, it is beyond the scope of this article, and may be revisited in a future publication.

The drift coefficient is then determined by fixing the time-asymptotic equilibrium distribution of the system.
For this purpose, we switch to the distribution-function representation of our problem in the subsequent section.

\subsection{Probability density function}

Choosing an SDE's time-evolution parameter~\(t\) singles out a preferred class of temporal hypersurfaces in Minkowskian space \cite{DunkelEtAl2009NP5}, which implicitly enter into the definition of the associated probability density function (PDF).
For \(\func{\tup{X}[0]}{t} \coloneq t + \const\) as in \cref{eq:SDE-position}, these hypersurfaces are isochronous hyperplanes, which corresponds to a simultaneous observation of the system in the laboratory frame -- a choice we deem appropriate for experimental measurements in spatially-extended particle detectors.
Together with \cref{eq:SDE-mass-shell}, this implies a factorization of the PDF for the \(4 + 4\) relativistic position and momentum coordinates,
\begin{equation}
	\func{\pdf{\func{\tupdelims{\tup{X}[0],\tup{X},\tup{P}[0],\tup{P}}}{t}}}{\tup{x}[0],\tup{x},\tup{p}[0],\tup{p}} = \func{\delta}{\tup{x}[0] - t} \x \func[\big]{\delta}{\textstyle \parens{\tup{p}[0]}^2 - E^2} \x \func{g}{t;\tup{x},\tup{p}}
	\mathinner{,}
\end{equation}
with \(E = \sqrt{m^2 + \sum_{i=1}^3 \parens{\tup{p}[i]}^2}\) and some
function~\(g\) of the time~\(t\) and \(3 + 3\) phase-space coordinates~\(\tupdelims{\tup{x},\tup{p}}\).
Integrating out \(\tup{x}[0]\) and \(\tup{p}[0]\) then yields \(\pdf{\func{\tupdelims{\tup{X},\tup{P}}}{t}} = \func{g}{t;\tup{x},\tup{p}} / \parens{2 E}\) for the phase-space PDF.
Here, the quantity \(\func{\pdf{\func{\tupdelims{\tup{X},\tup{P}}}{t}}}{\tup{x},\tup{p}} \x \diff{^3 x} \x \diff{^3 p}\) gives the probability that at time~\(t\) the Brownian particle is located at some position \(\tup{x} \in \prod_{i=1}^{3} \bracks{\tup{x}[i],\tup{x}[i] + \diff{\tup{x}[i]}}\) and carries momentum \(\tup{p} \in \prod_{i=1}^{3} \bracks{\tup{p}[i],\tup{p}[i] + \diff{\tup{p}[i]}}\) in the laboratory frame.

The SDE of any stochastic drift--diffusion process is intimately connected with the time evolution of its PDF.
The latter can be obtained by performing a Kramers--Moyal expansion, which yields a Fokker--Planck equation (FPE).
For the phase-space PDF, it takes the form
\begin{equation}
	\label{eq:FPE-position-momentum}
	\bracks[\big]{\derivop{t} - \func{\fpo{\tupdelims{\tup{X},\tup{P}}}}{\tup{x},\tup{p}}} \x \func{\pdf{\func{\tupdelims{\tup{X},\tup{P}}}{t}}}{\tup{x},\tup{p}} = 0
\end{equation}
with the (linear) Fokker--Planck operator (FPO)
\begin{equation}
	\label{eq:FPO-position-momentum}
	\func{\fpo{\tupdelims{\tup{X},\tup{P}}}}{\tup{x},\tup{p}}
	= - \sum_{i=1}^{3} \frac{\tup{p}[i]}{E} \x \derivop{\tup{x}[i]}
	+ \underbrace{\sum_{i=1}^{3} \derivop{\tup{p}[i]} \bracks[\bigg]{- \func{\drift[i]{\tup{P}}}{\tup{p}} + \tfrac{1}{2} \x \sum_{l=1}^{3} \func{\diffusion[i l]{\tup{P}}}{\tup{p}} \x \sum_{k=1}^{3} \derivop{\tup{p}[k]} \x \func{\diffusion[k l]{\tup{P}}}{\tup{p}}}}_{\func{\fpo{\tup{P}}}{\tup{p}}}
	\mathinner{.}
\end{equation}
Here, the second term is identical to the FPO governing the time-evolution of the momentum-space PDF
\begin{equation}
	\func{\pdf{\func{\tup{P}}{t}}}{\tup{p}}
	= \frac{1}{2 E} \x \int_{\mathbb{V}} \diff{^3 x} \x \func{g}{t;\tup{x},\tup{p}}
	\mathinner{,}
\end{equation}
which can easily be seen by integrating \cref{eq:FPE-position-momentum} over the set~\(\mathbb{V}\) of all possible particle positions.

To obtain an FPE for the rapidity-space PDF
\begin{equation}
	\func{\pdf{\func{\Rap}{t}}}{\rap} = \pi \x \pT \x \mT \x \int_{\mathbb{V}} \diff{^3 x} \x \func[\big]{g}{t;\tup{x},m \x \func{\sinh}{h},0,m \x \func{\cosh}{h} \x \func{\sinh}{y}}
\end{equation}
with \(\rap \coloneq \tupdelims{h,y}\), one can either transform the momentum-space FPE or perform a Kramers--Moyal expansion of the rapidity-space SDE~\labelcref{eq:SDE-rapidity}.
Both approaches yield the same equation
\begin{equation}
	\label{eq:FPE-rapidity}
	\bracks[\big]{\derivop{t} - \func{\fpo{\Rap}}{\rap}} \x \func{\pdf{\func{\Rap}{t}}}{\rap} = 0
\end{equation}
with the rapidity-space FPO
\begin{equation}
	\label{eq:FPO-rapidity}
	\func{\fpo{\Rap}}{\rap} = \sum_{i=1}^{2} \derivop{\rap[i]} \bracks[\bigg]{- \func{\drift[i]{\Rap}}{\rap} + \diffusivity[i k]{\Rap} \x \sum_{k=1}^{2} \derivop{\rap[k]}}
	\mathinner{.}
\end{equation}
Here, we assumed \(\diffusion{\Rap}\) to be constant with respect to \(\rap\), and combined the two diffusion coefficients into the (\(2 \times 2\))-dimensional diffusivity of \(\Rap\), which is given by
\begin{equation}
	\diffusivity[i k]{\Rap} \coloneq \tfrac{1}{2} \x \sum_{l=1}^{2} \diffusion[i l]{\Rap} \x \diffusion[k l]{\Rap}
	\quad\text{for}\quad
	i,k = 1,2
	\mathinner{.}
\end{equation}

\Cref{eq:FPE-rapidity} can be rewritten as a continuity equation with advective flux \(\func{\tup{j}_{\mathrm{adv}}}{\rap} = \func{\drift{\Rap}}{\rap} \x \func{\pdf{\func{\Rap}{t}}}{\rap}\) and diffusive flux \(\func{\tup{j}_{\mathrm{dif}}}{\rap} = - \diffusivity{\Rap} \tupderivop{\rap} \func{\pdf{\func{\Rap}{t}}}{\rap}\), offering a more intuitive perspective on the rather abstract drift and diffusion coefficients introduced in the preceding section.

The PDF formulation now offers a way to connect the functional form of drift and diffusion coefficient by assuming that the system is in detailed balance.
This ensures that the particle distribution approaches an equilibrium state \cite{nonthermal_equilibrium}, \(\lim_{t \to \infty} \func{\pdf{\func{\Rap}{t}}}{\rap} \equiv \func{\pdf{\Rap_\equ}}{\rap}\), so that drift and diffusivity fulfill the fluctuation--dissipation relation
\begin{equation}
	\label{eq:FDR}
	\func{\drift{\Rap}}{\rap} = - \diffusivity{\Rap} \x \tupderivop{\rap} \func{\potential{\Rap}}{\rap}
\end{equation}
with the generalized potential
\begin{equation}
	\func{\potential{\Rap}}{\rap} \coloneq - \func[\big]{\ln}{\func{\pdf{\Rap_\equ}}{\rap}}
	\mathinner{.}
\end{equation}

To reduce both mathematical complexity and the number of model parameters, we further assume \(\diffusion[1 1]{\Rap} \x \diffusion[2 1]{\Rap} + \diffusion[1 2]{\Rap} \x \diffusion[2 2]{\Rap} = 0\) in the following, so that the diffusivity matrix is diagonal, \(\diffusivity{\Rap} \eqcolon \func{\operatorname{diag}}{\DT,\DL}\).
The rapidity-space FPO then decomposes into a purely transverse and longitudinal part,
\begin{equation}
	\func{\fpo{\Rap}}{\rap}
	= \DT \x \derivop{h} \bracks*{\deriv{\func{\potential{\Rap}}{\rap}}{h} + \derivop{h}}
	+ \DL \x \derivop{y} \bracks*{\deriv{\func{\potential{\Rap}}{\rap}}{y} + \derivop{y}}
	\mathinner{.}
\end{equation}

As an alternative to our approach, a similarly simple equation could be constructed in momentum space by assuming a constant diagonal diffusivity for \(\tupdelims{\PT,\PL}\); however, the numerical solution of the time evolution would then be considerably more intricate:
Solving the associated FPE in momentum space would be challenging due to the PDFs' slow decline at high momenta, and transforming it to rapidity space would result in a diffusivity that is nonconstant and nondiagonal.
Assuming a constant diagonal diffusivity for \(\tupdelims{\PT,Y}\) instead would mitigate these problems, but we find it difficult to justify such a choice for this rather disparate pair of coordinates.

\subsection{Equilibrium state}

Instead of deriving the drift coefficient from microscopic considerations, we use an FDR of the form~\labelcref{eq:FDR} to determine it as a function of the diffusivity and the equilibrium state of the system.
If the interactions between the particles produced in a relativistic heavy-ion collision did not cease due to physical freeze-out but had the chance to evolve undisturbed for an extended period of time, it can be expected that the system would thermalize and reach a stationary state in momentum space.
In position space, however, the system should continue to expand into the surrounding vacuum due to a lack of spatial confinement, resulting in a finite collective particle flow.

Therefore, we model the equilibrium state of any produced charged hadron in momentum space by means of a generalized Maxwell--Jüttner distribution for an expanding thermal reservoir,
\begin{equation}
	\label{eq:pdf-thermal}
	\func{\pdf{\tup{P}_\equ}}{\tup{p}} \coloneq \frac{C_{\mathrm{th}}}{V} \x \bracks*{\int_{\Sigma} \frac{\diff{\tup*{\sigma}[\mu]} \x \tup{p}[\mu]}{\tup{p}[0]} \x \func*{\exp}{\frac{m - \tup*{u}[\nu] \x \tup{p}[\nu]}{T}}}_{\tup{p}[0] = E}
	\mathinner{,}
\end{equation}
where summation over pairs of (Greek) indices is implied with metric signature~\(\tupdelims{+,-,-,-}\).

The symbol~\(\Sigma\) denotes the three-dimensional spacetime hypersurface containing the reservoir and \(\tup{u}[\mu]\) its proper collective-flow velocity, which may vary within \(\Sigma\).
Consequently, the expression \(V \coloneq \int_{\Sigma} \diff{\tup*{\sigma}[\mu]} \x \tup{u}[\mu]\) corresponds to the volume of the expanding reservoir.
The remaining normalizing constant~\(C_{\mathrm{th}}\) can be calculated analytically by substitution of variables with the rapidity coordinates defined in \cref{eq:rapidity-coordinates}, yielding
\begin{equation}
	C_{\mathrm{th}} = \frac{\kappa}{4 \pi m^3 \x \func{\exp}{\kappa} \x \func{K_2}{\kappa}}
\end{equation}
independent of \(\Sigma\) and \(\tup{u}[\mu]\).
\(K_2\) is the modified Bessel function of the second kind and order \(2\), while \(\kappa \coloneq m / T\) is used as a shorthand for the dimensionless ratio of the particle's rest and thermal energy.

A concrete implementation of \cref{eq:pdf-thermal}
--~which includes the (nongeneralized) Maxwell--Jüttner distribution~--
that quickly comes to mind is a purely spherical flow with kinetic freeze-out at constant local time \cite{LeeEtAl1990ZPC48}.
As this family of distributions tends to broaden with increasing temperature, it is instructive to study their infinite-temperature limit (\(\kappa \to 0\)), which then provides an upper bound for the PDF's width.
It can easily be shown that in longitudinal-rapidity space, the marginal PDF of \(Y_\equ\) approaches the function \(1 / \bracks{2 \func{\cosh}{y}^2}\), implying that its full width at half maximum (FWHM) cannot exceed \(2 \func{\arcosh}{\sqrt{2}}\) or, equivalently, its variance is bound by \(\pi^2 / 12\) at all temperatures.
Comparing this threshold with available LHC charged-hadron data quickly reveals that this is significantly too narrow to provide a sensible model for the experimentally observed distributions.

Therefore, we use a cylindrical-shaped expansion that spreads faster in the longitudinal than in the transverse direction.
The result can be understood as a Bjorken flow \cite{Bjorken1983PRD27} with maximum longitudinal flow rapidity~\(\ysurf > 0\), \cite{SchnedermannEtAl1993PRC48}
\begin{subequations}
\label{eq:pdf-equ}
\begin{equation}
	\label{eq:pdf-equ-convolution}
	\func{\pdf{\Rap_\equ}}{\rap} \coloneq \frac{1}{2 \ysurf} \x \int_{-\ysurf}^{+\ysurf} \diff{y'} \x \func{\pdf{\Rap_\equ,\trv}}{h,y - y'}
	\mathinner{,}
\end{equation}
on top of a purely transverse expansion,
\begin{multline}
	\label{eq:pdf-equ-kernel}
	\func{\pdf{\Rap_\equ,\trv}}{\rap} \coloneq \frac{2 \pi m^3 C_{\mathrm{th}}}{\func*{\hgf}{\tfrac{1}{2},\tfrac{1}{\nu};1 + \tfrac{1}{\nu};\betasurf^2}} \x \func{\sinh}{h} \x \func{\cosh}{h}^2 \x \func{\cosh}{y}
	\times \\ \times
	\frac{2}{R_{\trv}^2} \x \int_0^{R_{\trv}} \diff{r_{\trv}} \x r_{\trv} \x \func[\big]{I}{\kappa - \kappa \x \sqrt{1 + \abs{\vec{u}_{\trv}}^2} \x \func{\cosh}{h} \x \func{\cosh}{y},\kappa \x \abs{\vec{u}_{\trv}} \x \func{\sinh}{h}}
	\mathinner{.}
\end{multline}
\end{subequations}
For the transverse velocity profile~\(\abs{\vec{u}_{\trv}}\), a power law of the form
\begin{equation}
	\frac{\abs{\vec{u}_{\trv}}}{\sqrt{1 + \abs{\vec{u}_{\trv}}^2}} \overset{!}{=} \parens*{\frac{r_{\trv}}{R_{\trv}}}^\nu \x \betasurf
\end{equation}
is assumed, where \(\betasurf > 0\) agrees with the reservoir's (nonrelativistic) surface velocity at \(y = 0\).
Based on analyses of LHC transverse-momentum data \cite{ALICE2013PRC88,ALICE2020PRC101}, we choose \(\nu = \num{0.7}\) for the power-law exponent.
The auxiliary function
\begin{equation}
	\func{I}{a_1,a_2} \coloneq \func{\exp}{a_1} \x \func{I_0}{a_2}
\end{equation}
--~the intent behind its introduction will become clear in a moment~--
encapsulates an exponential and a modified Bessel function of the first kind and order \(0\).

It is well-known \cite{MichaelVanryckeghem1977JPGNP3,Hagedorn1983RNC6} that conventional thermal models fail to describe the transverse high-momentum tails of charged-hadron data, which do not follow an exponential but a Pareto distribution (\enquote{power law}).
This is often attributed to the increasing importance of \enquote{hard} processes in this regime, which are not covered by thermal physics.
In this work, this observed transition shall be accounted for phenomenologically, as inspired by the parameterization introduced by Hagedorn \cite{Hagedorn1983RNC6}.
We replace the exponential function in \cref{eq:pdf-thermal} by a member of the exponential sequence \cite{exponential_sequence},
\begin{equation}
	\func{\widetilde{\exp}}{x;n} \coloneq \parens*{1 + \frac{x}{n}}^n \approx \begin{cases}
		\func{\exp}{x} & \text{for} \quad \abs*{\frac{x}{n}} \ll 1
		\\
		\parens*{\frac{x}{n}}^n & \text{for} \quad \abs*{\frac{x}{n}} \gg 1
	\end{cases}
\end{equation}
with \(n < 0\),
which exhibits the required asymptotic behavior for its argument
\begin{equation}
	x \coloneq \frac{m - \tup*{u}[\nu] \x \tup{p}[\nu]}{T}
	\mathinner{.}
\end{equation}
The inclusion of the particle's mass is crucial here to reproduce the nonrelativistic limit
--~a normal distribution~--
for low momenta.
This approach offers some advantages over a piece-wise definition, as it allows a smooth transition and introduces only one additional parameter.
Moreover, it naturally incorporates the longitudinal degrees of freedom and generalizes well within our family of thermal distributions with collective flow.

For the cylindrical-expanding distribution discussed above, replacing the exponential and performing the integral amounts to substituting \(I\) in \cref{eq:pdf-equ-kernel} with the function
\begin{equation}
	\func{\tilde{I}}{a_1,a_2;n} = \func{\widetilde{\exp}}{a_1;n} \x \parens*{1 - \tfrac{a_2}{n + a_1}}^n \x \func*{\hgf}{-n,\tfrac{1}{2};1;\tfrac{2 a_2}{a_2 - a_1 - n}}
\end{equation}
involving the ordinary hypergeometric function~\(\hgf\); sending \(n \to - \infty\) recovers the original definition.
The modified distribution functions remain normalizable for \(n < -3\), which constitutes an upper bound for the exponent.
However, the normalizing factor~\(\tilde{C}_{\mathrm{th}}\) must now be computed numerically.

\begin{figure}
	\centering
	\includegraphics[scale=0.85]{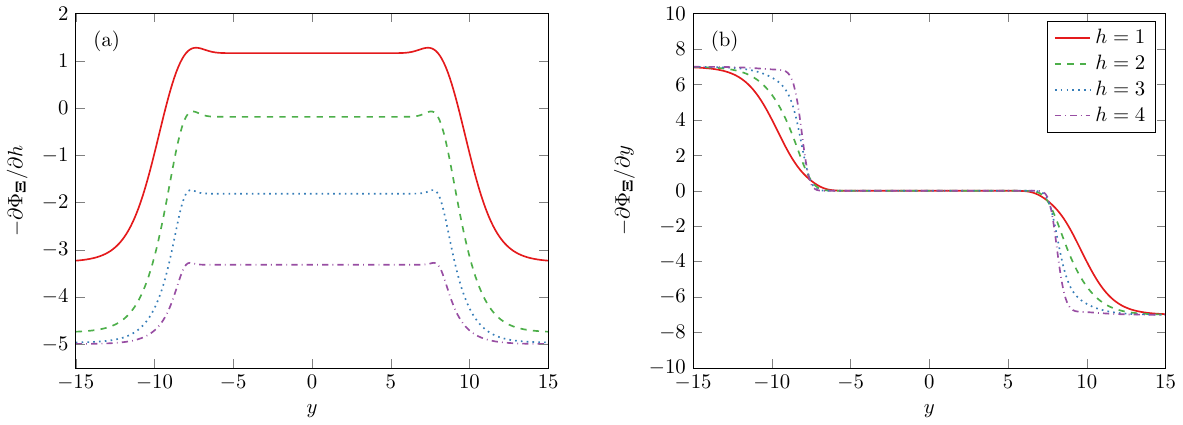}
	\caption{
		Negative gradient of the generalized potential~\(\potential{\Rap}\) in rapidity space for the (a) transverse and (b) longitudinal direction as a function of transverse (\(h\)) and longitudinal rapidity (\(y\)).
		The parameters (see text) are chosen to roughly reflect the effect of a \qty{2.76}{\TeV} Pb--Pb collision (\(\ysurf = \num{8}\), \(\betasurf = \num{0.64}\), \(\nu = \num{0.7}\), \(n = -8\)) on charged pions (\(\kappa = \num{1.4}\)); see \cref{sec:results} for more details on our choice of parameters.
		For these particles, the four transverse-rapidity cuts~\(h = \numlist{1;2;3;4}\) correspond to \(\pT \approx \qtylist{0.16;0.51;1.4;3.8}{\GeV}\), respectively.
	}
	\label{fig:1}
\end{figure}

A welcome side effect of this approach is that the gradient of the generalized potential, \(\tupderivop{\rap} \func{\potential{\Rap}}{\rap}\), remains finite for all \(\rap\) with \(h > 0\), which benefits stability when solving the FPE numerically.
Examination of the high-rapidity regime reveals that the transverse component cannot exceed \(- \parens{n + 3}\), while the longitudinal component is bounded from above and below by \(\mp \parens{n + 1}\), respectively.
This is shown exemplarily in \cref{fig:1} for a parameter set that roughly emulates a \qty{2.76}{\TeV} Pb--Pb collision; more details on our choice of parameters are given in \cref{sec:results}.
Here, we show the components of the negative gradient, which generate the deterministic generalized forces~\(\drift{\Rap}\) when multiplied with the diffusivity~\(\diffusivity{\Rap}\); see \cref{eq:FDR}.
In our case of a diagonal diffusivity, the negative components are simply identical to \(\drift{\Rap}\) up to scaling with \(\DT\) and \(\DL\), respectively, which makes interpretation of the plot relatively straightforward:
Longitudinally, particles with \(\abs{y} > \ysurf\) are moved towards midrapidity, where the deterministic forces vanish so that dynamics is mainly governed by the remaining stochastic forces.
In the transverse direction, particles with high \(h\) are moved towards \(h = 0\) for all \(y\), while a reversal of the direction occurs for lower \(h\) when crossing \(\pm \ysurf\).
Although the magnitude of the transverse forces diverges for \(h \to 0\), this is not a problem in practice since the particle density vanishes simultaneously.

The clearly visible plateaus in the region \(\abs{y} \leq \ysurf\) are a consequence of the fact that \cref{eq:pdf-equ-convolution} is effectively a convolution of \cref{eq:pdf-equ-kernel} with a rectangular pulse.
This is an artifact of the current simplified treatment of longitudinal flow and could be improved in the future.

An analysis of ALICE p--p data \cite{ALICE2011EPJC71} suggests that the exponent is close to \num{-8} at the LHC, which we will use as an approximation for \(n\) in the following.

\subsection{Initial state}

We assume the thermalization process to begin directly after the production of particles from hadronic interactions of the colliding nuclei, and model the system's time evolution starting from this initial time~\(t_\ini\).
For the associated initial particle distributions, we use a QCD-inspired phenomenological framework based on gluon saturation in the deep-inelastic scattering of participant partons (specifically gluons and valence quarks) \cite{Gribov1983PR100,MuellerQiu1986NPB268,BlaizotMueller1987NPB289,McLerranVenugopalan1994PRD49}, which we could already successfully employ in our previous studies on baryon stopping \cite{HoelckWolschin2020PRR2}.

A fundamental assumption within this framework is that the gluon density saturates below a characteristic momentum scale~\(\Qs\), so that the respective gluons form a color-glass condensate (CGC).
This gluon saturation scale can be parameterized as \cite{Golec-BiernatWuesthoff1998PRD59}
\begin{equation}
	\func{\Qs^2}{x} = \sqrt[3]{A} \x Q_0^2 \x \parens*{\frac{x_0}{x}}^\lambda
\end{equation}
for small values of the gluon's Bjorken momentum fraction~\(x\), where we use \(\lambda = \num{0.2}\) and \(\tfrac{4}{9} Q_0^2 \x x_0^\lambda = \qty{0.09}{\GeV\squared}\) to be consistent with our prior baryon-stopping analysis \cite{HoelckWolschin2020PRR2}.
These values can be compared with literature results, where \(\lambda \approx \num{0.288}\) and \(Q_0^2 \x x_0^\lambda \approx \qty{0.097}{\GeV\squared}\) were determined in a fit to deep-inelastic scattering e--p data from HERA \cite{Golec-BiernatWuesthoff1998PRD59}.
The interaction of the participating nucleons is assumed to be mediated by quark--antiquark pairs (\enquote{dipoles}) emitted by the confined partons, which then inelastically scatter off the partons of the oppositely moving nucleus.
Subsequent recombination of the involved quarks and gluons results in the production of new hadrons, which are emitted from the nuclear fragments.

If the transverse and longitudinal rapidity \(h\) and \(y\) as well as the mass~\(m\) of the produced hadron are known, the Bjorken momentum fractions of the two generating partons can be reconstructed from kinetic considerations as
\begin{equation}
	x_\pm \approx \frac{m \x \func{\cosh}{h} \x \func{\exp}{\pm y}}{2 \mN \x \func{\sinh}{\yb}}
	\mathinner{,}
\end{equation}
where \(\mN\) is the average nucleon mass of the colliding nuclei and \(x_+\) (\(x_-\)) denotes the Bjorken \(x\) of the parton contained in the forward-moving (backward-moving) nucleon.

At low Bjorken \(x\), the nucleon momentum is mostly carried by gluons, while valence quarks take the main share at high \(x\).
For two interacting partons, this results in four dominant interaction processes.
The bulk of produced charged hadrons originates from gluon--gluon interactions (\(\gluglu\)), which predominantly yield hadronic particle--antiparticle pairs with low to intermediate \(y\).
Their distribution in rapidity space can be approximated with \cite{KharzeevEtAl2005NPA747}
\begin{equation}
	\label{eq:pdf-ini-gg}
	\func{\pdf{\func{\Rap_\gluglu}{t_\ini}}}{\rap} \approx C_{\gluglu} \x \func{\sinh}{h} \x \func{\cosh}{h} \x \prod_{i \in \setdelims{+,-}} x_i \x \frac{\func[\big]{G}{x_i;m \x \func{\sinh}{h}}}{\func{\sinh}{h}^2} \x \func{\Theta}{1 - x_i}
	\mathinner{,}
\end{equation}
which is essentially the product of two simplified gluon structure functions
\begin{equation}
	x \x \func{G}{x;Q} \propto \parens{1 - x}^4 \x \func[\big]{\min}{Q^2,\func{\Qs^2}{x}}
	\mathinner{.}
\end{equation}
The next largest source of charged-particle production are interactions of gluons and valence quarks.
For these particles, we employ a simplified distribution that is obtained from multiplication of the parton distribution function~\(q_{\mathrm{v}}\) of the valence quarks \cite{MartinEtAl2002PLB531} and their dipole scattering amplitude with the opposing gluons \cite{DumitruEtAl2006NPA765}
\begin{equation}
	\func{\varphi}{x;Q} \approx \frac{4 \pi}{\tfrac{4}{9} \func{\Qs^2}{x}} \x \func*{\exp}{- \frac{Q^2}{\tfrac{4}{9} \func{\Qs^2}{x}}} \x \parens{1 - x}^4
	\mathinner{.}
\end{equation}
For a forward-moving valence quark and a backward-moving gluon (\(\quaglu\)), the resulting simplified PDF reads \cite{KharzeevEtAl2004PLB599,BaierEtAl2006NPA764,DumitruEtAl2006NPA765}
\begin{equation}
	\label{eq:pdf-ini-qg}
	\func{\pdf{\func{\Rap_\quaglu}{t_\ini}}}{\rap} \approx C_{\quaglu} \x \func{\sinh}{h} \x \func{\cosh}{h} \x x_+ \x \func{q_{\mathrm{v}}}{x_+} \x \func{\Theta}{1 - x_+} \x \func[\big]{\varphi}{x_-;m \x \func{\sinh}{h}} \x \func{\Theta}{1 - x_-}
	\mathinner{.}
\end{equation}
The PDF for the inverse process \(\gluqua\) follows directly from the interchange of \(x_+\) and \(x_-\),
\begin{equation}
	\label{eq:pdf-ini-gq}
	\func{\pdf{\func{\Rap_\gluqua}{t_\ini}}}{h,y} = \func{\pdf{\func{\Rap_\quaglu}{t_\ini}}}{h,- y}
	\mathinner{.}
\end{equation}
All normalizing constants (\(C_{\gluglu}\), \(C_{\quaglu} = C_{\gluqua}\)) are determined numerically.

The contribution from interactions of valence quarks with valence quarks
is expected to be small compared to the processes mentioned above and will be neglected here.
For the remaining three sources of charged-hadron production, we construct independent drift--diffusion processes whose initial states are given by \cref{eq:pdf-ini-gg,eq:pdf-ini-qg,eq:pdf-ini-gq}, respectively.

\section{Results}
\label{sec:results}

When solving the FPE~\labelcref{eq:FPE-rapidity} numerically for \(t \in \bracks{t_\ini,t_\fin}\), it is convenient to perform the variable substitution
\begin{equation}
	t \mapsto \delta \coloneq \parens{t - t_{\ini}} / \incr{t} \in \bracks{0,1}
\end{equation}
with the interaction duration~\(\incr{t} \coloneq t_{\fin} - t_{\ini}\).
The transformed FPE is physically dimensionless, and contains the diffusivity solely as a product with the interaction duration, \(\diffusivity{\Rap} \x \incr{t}\).
This reduces our number of free model parameters by one, as both \(\diffusivity{\Rap}\) and \(\incr{t}\) are experimentally inaccessible to date.

The majority of charged particles produced at the LHC are pions, followed by kaons and (anti)protons, where the measured particle ratios \cite{ALICE2013PRC88,ALICE2020PRC101} are approximately \(R_{\cpion} : R_{\ckaon} : R_{\cnucleon} \approx \qty{83}{\percent} : \qty{13}{\percent} : \qty{4}{\percent}\).
From our numerical solutions of the above Fokker--Planck equation, we thus construct the marginal pseudorapidity spectrum of charged hadrons as
\begin{equation}
	\label{eq:ndf-L}
	\frac{\diff{N}}{\diff{\eta}} \overset{!}{=} \mkern-12mu \sum_{\crampedsubstack{j = \gluglu,\quaglu,\gluqua \\ k = \cpion,\ckaon,\cnucleon}} \mkern-12mu N_j \x R_k \x \int_0^\infty \diff{h} \x \abs*{\deriv{\tupdelims{h,y}}{\tupdelims{h,\eta}}} \x \func{\pdf{\func{\Rap_{j \to k}}{t_\fin}}}{\rap}
	\mathinner{,}
\end{equation}
where the transformation from transverse and longitudinal rapidity to pseudorapidity
\begin{equation}
	\eta \coloneq \func*{\artanh}{\frac{\pL}{\abs{\vec{p}}}} = \func*{\arsinh}{\sqrt{1 + \func{\sinh}{h}^{-2}} \x \func{\sinh}{y}}
\end{equation}
is mediated by the Jacobian
\begin{equation}
	\abs*{\deriv{\tupdelims{h,y}}{\tupdelims{h,\eta}}} = \frac{1}{\sqrt{1 + \func{\sinh}{h}^{-2} \x \func{\cosh}{\eta}^{-2}}}
	\mathinner{.}
\end{equation}
Allocation of the total charged-particle number to the three particle production processes considered is done via the free parameters \(N_{\gluglu} > N_{\quaglu} = N_{\gluqua}\).

Due to the two-dimensional nature of our model, the joint transverse-momentum and pseudorapidity spectrum is also readily available and can be obtained via
\begin{equation}
	\label{eq:ndf-TL}
	\frac{\diff{^2 N}}{\diff{\pT} \x \diff{\eta}} \overset{!}{=} \mkern-12mu \sum_{\crampedsubstack{j = \gluglu,\quaglu,\gluqua \\ k = \cpion,\ckaon,\cnucleon}} \mkern-12mu N_j \x R_k \x \abs*{\deriv{\tupdelims{h,y}}{\tupdelims{\pT,\eta}}} \x \func{\pdf{\func{\Rap_{j \to k}}{t_\fin}}}{\rap}
\end{equation}
with the corresponding Jacobian
\begin{equation}
	\abs*{\deriv{\tupdelims{h,y}}{\tupdelims{\pT,\eta}}} = \frac{1}{\sqrt{m^2 + \pT^2} \x \sqrt{1 + \parens*{\tfrac{m}{\pT \x \func{\cosh}{\eta}}}^2}}
	\mathinner{.}
\end{equation}

\bigbreak

We compare our results to corresponding data from ALICE \cite{ALICE2013PLB726,ALICE2017PLB772,ALICE2018JHEP2018} and ATLAS \cite{ATLAS2015JHEP2015} for central Pb--Pb collisions at LHC energies \(\sqrtsNN = \qtylist{2.76;5.02}{\TeV}\).
In our analysis, we concentrate on the datapoints not smaller than \qty{5}{\percent} of the data's peak value, as the precision of our numerical solutions is not sufficient across multiple orders of magnitude.
This effectively confines our results and analyses to the region \(\pT \leq \qty{2}{\GeV}\).

Apart from the source-dependent particle numbers (\(N_{\gluglu}\), \(N_{\quaglu} = N_{\gluqua}\)), free parameters of the model are the two diffusivity matrix elements times interaction duration (\(\DT \x \incr{t}\), \(\DL \x \incr{t}\)), the two flow parameters \(\ysurf\) and \(\betasurf\), and the equilibrium temperature~\(T\), which we restrict to the interval \(\bracks{\qty{100}{\MeV},\qty{160}{\MeV}}\).
They are determined through weighted least-square fits of our calculated charged-hadron distribution functions~\labelcref{eq:ndf-L,eq:ndf-TL} to the experimental data.
For each datapoint, we average our calculated function over the associated coordinate bin, subtract the result from the respective measured particle number, and then divide by the total (statistical plus systematic) experimental uncertainty.
The obtained normalized residuals are squared and added, where we sum over all datapoints with the same center-of-mass energy to achieve the best possible phase-space coverage.
Minimizing the resulting \(\chi^2\)s yields the parameter estimates listed in \cref{tab:1}.

As an intermediate result of our fits, we found the transverse degrees of freedom to thermalize significantly faster than the longitudinal ones, \(\DT \x \incr{t} \gg \DL \x \incr{t}\).
While this meets physical expectation, the FPE tends to become numerically unstable in this regime.
Therefore, we adapted our solution algorithm to impose a nearly instantaneous thermalization in the transverse direction, which corresponds to the limit \(\DT \x \incr{t} \to \infty\) and recovers numerical stability.
Details on our approach are given in \cref{app:fast-transverse-equilibration}.

Also, using the same maximum longitudinal flow rapidity and longitudinal diffusivity times interaction duration for all sources did not provide a good description of the data.
In our fits, we hence decoupled the \(\gluglu\) source from the \(\quaglu\) and \(\gluqua\) sources in this respect.
Since the \(\gluglu\) flow rapidity always came to lie close to the beam rapidity, we identified \(\parens{\ysurf}_{\gluglu} = \yb\), which fits well with physical intuition.

In a prior investigation \cite{Wolschin2016PRC94}, it was found that the particle numbers in the central gluon--gluon source scale with the cubic logarithm of the center-of-mass energy.
This is confirmed in the present investigation when comparing our results at the two LHC energies shown in \cref{tab:1}:
The ratio of our \(N_{\gluglu}\) estimates is \num{0.786+-0.028}, whereas the cubic ratio of the beam rapidities --~which are proportional to the logarithms of the beam energies~-- is \num{0.805}, which is compatible within the margin of error.

\begin{table}
	\centering
	\caption{
		Free model parameters obtained from weighted least-square fits to experimental data from ALICE \cite{ALICE2013PLB726,ALICE2017PLB772,ALICE2018JHEP2018} and ATLAS \cite{ATLAS2015JHEP2015}; uncertainty estimates are based on the covariance matrix in free-parameter space.
		Starred values at \qty{5.02}{\TeV} were extrapolated from the \qty{2.76}{\TeV} results (see text), and then treated as fixed parameters in the \qty{5.02}{\TeV} fit.
		For the equilibrium temperature~\(T\), both fits hit the lower interval boundary, and hence, no uncertainties are given for these parameters.
		The last column lists the goodness-of-fit in terms of the \(\chi^2\)-value divided by the number of degrees of freedom.
	}
	\label{tab:1}
	\renewcommand{\arraystretch}{1.25}
	\begin{tabular}[htbp]{cccccccccc}
		\toprule
		\(\sqrtsNN / \unit{\TeV}\) & \(\yb\)    & \(N_{\gluglu}\)  & \(N_{\quaglu}\) & \(\parens{\DL \x \incr{t}}_{\gluglu}\) & \(\parens{\DL \x \incr{t}}_{\quaglu}\) & \(\parens{\ysurf}_{\quaglu}\) & \(\betasurf\)        & \(T / \unit{\MeV}\) & \(\chi^2 / \mathrm{ndf}\)  \\
		\midrule
		\num{2.76}                 & \num{7.99} & \num{15860+-370} & \num{440+-120}  & \num{0.59+-0.47}                       & \num{0.769+-0.08}                      & \num{1.02+-0.14}              & \num{0.6405+-0.0040} & \num{100}           & \(\frac{\num{69.0}}{141}\) \\
		\num{5.02}                 & \num{8.58} & \num{20180+-550} & \num{420+-190}  & \num{0.63}*                            & \num{0.827}*                           & \num{0.79+-0.14}              & \num{0.691 +-0.052 } & \num{100}           & \(\frac{\num{86.7}}{ 56}\) \\
		\bottomrule
	\end{tabular}
\end{table}

\begin{figure}
	\centering
	\includegraphics[scale=0.85]{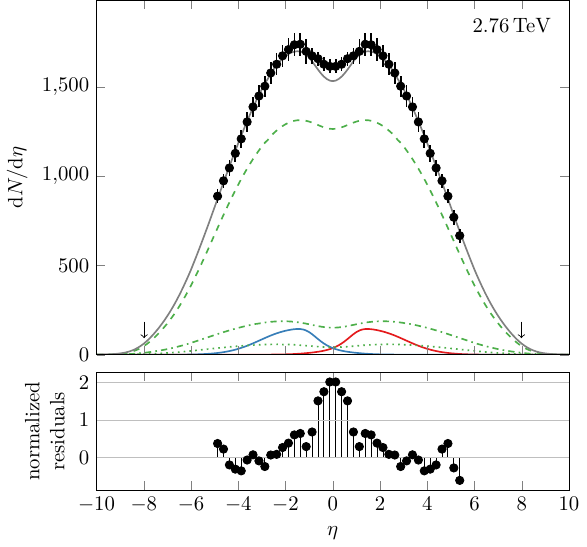}
	\caption{
		Calculated marginal pseudorapidity spectrum~\(\dNdeta\) of produced charged hadrons (upper solid curve, gray) in central \({\sqrtsNN = \qty{2.76}{\TeV}}\) Pb--Pb collisions, compared with \qtyrange{0}{5}{\percent} ALICE data \cite{ALICE2013PLB726}.
		Contributions from pions (dashed), kaons (dot-dashed), and (anti)protons (dotted) are shown for the central \(\gluglu\) source, together with the total yields from the forward \(\quaglu\) (red) and backward \(\gluqua\) (blue) fragmentation sources.
		Arrows indicate the beam rapidities, vertical bars the statistical plus systematic experimental uncertainties.
		The bins are smaller than the symbol size.
	}
	\label{fig:2}
\end{figure}
\begin{figure}
	\centering
	\includegraphics[scale=0.85]{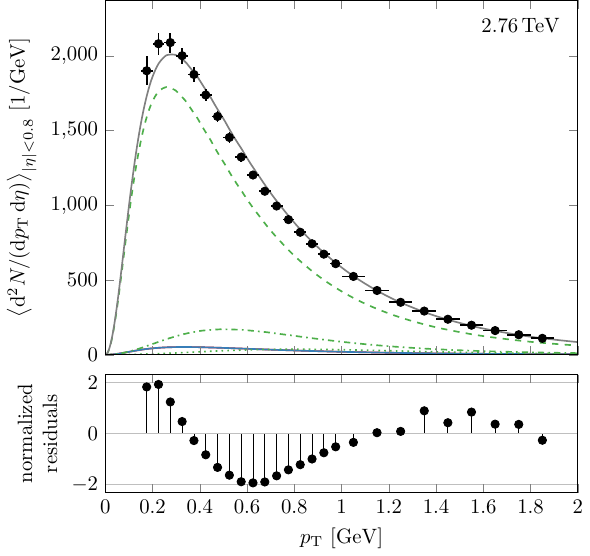}
	\caption{
		Calculated joint transverse-momentum and rapidity spectrum~\(\dNdpTdeta\) of produced charged hadrons (upper solid curve, gray)
		--~averaged for \({\abs{\eta} < \num{0.8}}\) and displayed as function of \(\pT\)~--
		in central \({\sqrtsNN = \qty{2.76}{\TeV}}\) Pb--Pb collisions, compared with \qtyrange{0}{5}{\percent} ALICE data \cite{ALICE2018JHEP2018}.
		Contributions from pions (dashed), kaons (dot-dashed), and (anti)protons (dotted) are shown for the central \(\gluglu\) source, together with the total yield from a single fragmentation source (lower solid curve, blue).
		Vertical bars indicate the statistical plus systematic experimental uncertainties, horizontal bars the bin sizes.
	}
	\label{fig:3}
\end{figure}

In \cref{fig:2}, calculated marginal rapidity spectra~\(\dNdeta\) of produced charged hadrons in central \qty{2.76}{\TeV} Pb--Pb collisions are compared with ALICE data \cite{ALICE2013PLB726}, upper frame, together with the normalized residuals (data minus calculation divided by uncertainty), lower frame.
Pions represent the largest contribution (dashed), about \qty{83}{\percent} of the total yield \cite{ALICE2013PRC88}, followed by kaons (about \qty{13}{\percent}, dot-dashed), and (anti)protons (about \qty{4}{\percent}, dotted).
The total contributions for pions, kaons, and protons from the forward (red) and backward (blue) fragmentation sources are shown as solid curves.
In the two-dimensional model, more stringent constraints are imposed on the size of the fragmentation sources than in the previous, purely one-dimensional phenomenological model \cite{Wolschin2013JPG40}, and the particle content of the fragmentation sources comes out to be smaller than in these earlier calculations.

While the transverse degrees of freedom are found to be very close to the equilibrium state at LHC energies (see above) the system is still far from the stationary state in its longitudinal degrees of freedom.
In the pseudorapidity spectrum, the calculated dip at midrapidity is mainly due to the Jacobian transformation from rapidity to pseudorapidity space, and to a lesser extent due to the effect of the fragmentation sources.
It is too deep in the model calculation -- a result that cannot be cured easily within the two-dimensional model due to the complex interplay of transverse and longitudinal degrees of freedom, while in effective one-dimensional models, the dip is often more pronounced due to the usage of an averaged Jacobian.
The origin of this deviation is most likely connected to our assumption of a constant diagonal diffusivity.
Another consequence of this assumption is the underprediction of the maximum in the corresponding joint spectrum that is shown in \cref{fig:3} as an averaged function of \(\pT\) for \(\abs{\eta} < \num{0.8}\) in comparison with \qty{2.76}{\TeV} ALICE data \cite{ALICE2018JHEP2018}.
Again, contributions from pions, kaons, and protons are shown, together with the (overlapping) yields from the forward and backward fragmentation sources.
Here, the latter are very small because the plot only covers the midrapidity region \(\abs{\eta} < \num{0.8}\), where the combined yield from the fragmentation sources has a local minimum, see \cref{fig:2}.

\begin{figure}
	\centering
	\includegraphics[scale=0.85]{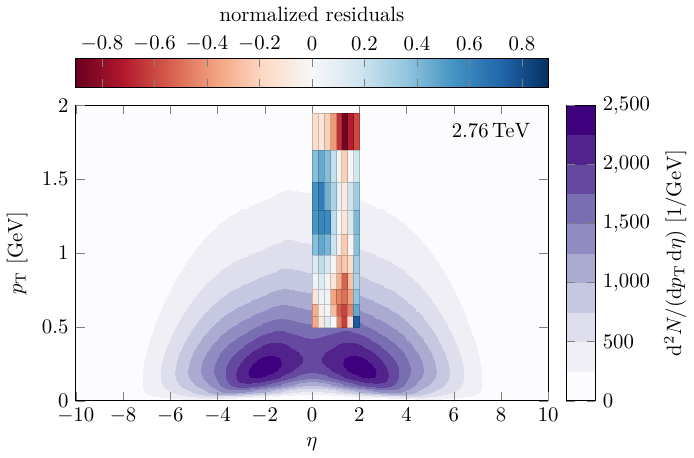}
	\caption{
		Calculated joint transverse-momentum and rapidity spectrum~\(\dNdpTdeta\) of produced charged hadrons (color scale to the right) in central \({\sqrtsNN = \qty{2.76}{\TeV}}\) Pb--Pb collisions, compared with \qtyrange{0}{5}{\percent} ATLAS data \cite{ATLAS2015JHEP2015}.
		Normalized residuals (data minus calculations divided by uncertainties) are indicated as rectangles for the respective bins (color scale at the top).
	}
	\label{fig:4}
\end{figure}
\begin{figure}
	\centering
	\includegraphics[scale=0.85]{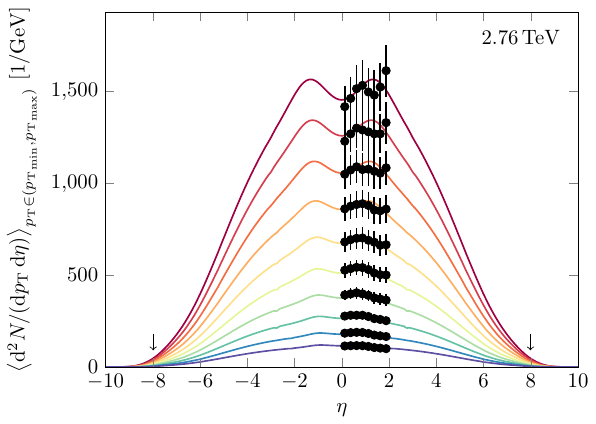}
	\caption{
		Calculated joint transverse-momentum and pseudorapidity spectrum~\(\dNdpTdeta\) of produced charged hadrons (solid curves)
		--~averaged over the individual experimental \(\pT\) bins (\qtyrange{0.5}{1.95}{\GeV}, top to bottom) and displayed as multiple functions of \(\eta\)~--
		in central \({\sqrtsNN = \qty{2.76}{\TeV}}\) Pb--Pb collisions, compared with \qtyrange{0}{5}{\percent} ATLAS data \cite{ATLAS2015JHEP2015}.
		Arrows indicate the beam rapidities, vertical bars the statistical plus systematic experimental uncertainties.
		The \(\eta\) bins are smaller than the symbol size.
	}
	\label{fig:5}
\end{figure}

The calculated joint spectrum~\(\dNdpTdeta\) of produced charged hadrons in \qty{2.76}{\TeV} Pb--Pb collisions is shown in full in \cref{fig:4} as a contour plot (scale on the right), with normalized residuals (measurement minus calculation divided by uncertainty) derived from ATLAS data \cite{ATLAS2015JHEP2015} (scale at the top).
\Cref{fig:5} depicts the same calculated spectrum and experimental data, but presented \(\pT\)-binwise
--~from \qtyrange{0.5}{1.95}{\GeV}, top to bottom~--
plotted against pseudorapidity.
Each of the curves shown is an average of the calculated spectrum with respect to one of the \(\pT\) bins.
Our calculated results agree with the ATLAS data (\(\num{0} < \eta < \num{2}\)) within the experimental error bars, albeit the rightmost datapoints at \(\eta \approx \num{2}\) stand out at low transverse momenta.

As could already be seen in \cref{fig:2,fig:4}, the charged-hadron yield at the beam rapidity (arrows) is substantially larger than zero, which is essentially a consequence of the Jacobian transformation from rapidity to pseudorapidity space, as well as the pion mass being considerably smaller than the masses of the colliding nucleons.

\begin{figure}
	\centering
	\includegraphics[scale=0.85]{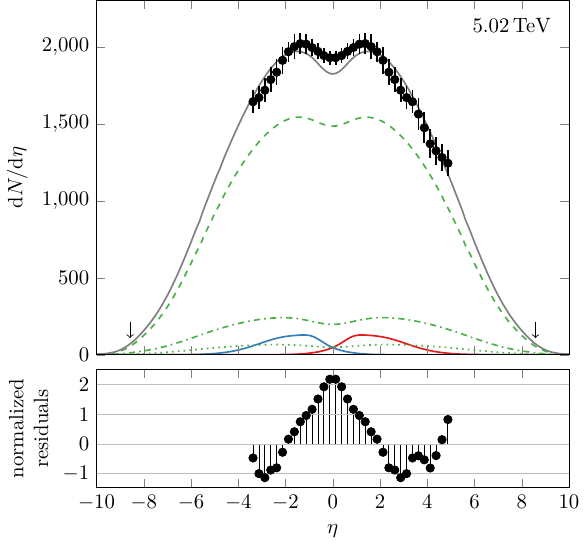}
	\caption{
		Calculated marginal pseudorapidity spectrum~\(\dNdeta\) of produced charged hadrons (upper solid curve, gray) in central \({\sqrtsNN = \qty{5.02}{\TeV}}\) Pb--Pb collisions, compared with \qtyrange{0}{5}{\percent} ALICE data \cite{ALICE2017PLB772}.
		Contributions from pions (dashed), kaons (dot-dashed), and (anti)protons (dotted) are shown for the central \(\gluglu\) source, together with the total yields from the forward \(\quaglu\) (red) and backward \(\gluqua\) (blue) fragmentation sources.
		Arrows indicate the beam rapidities, vertical bars the statistical plus systematic experimental uncertainties.
		The bins are smaller than the symbol size.
	}
	\label{fig:6}
\end{figure}
\begin{figure}
	\centering
	\includegraphics[scale=0.85]{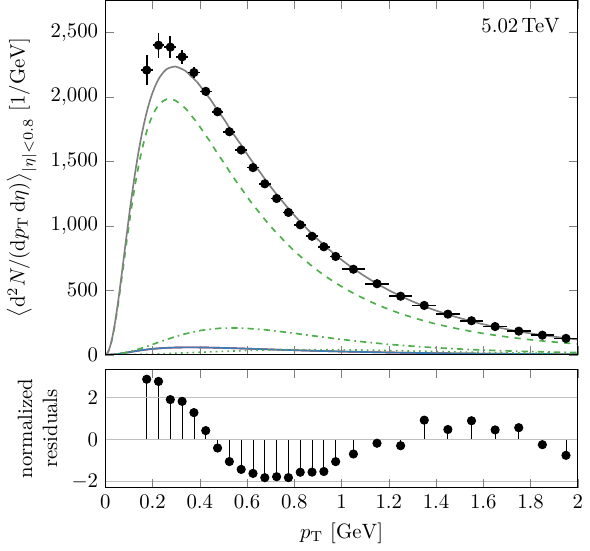}
	\caption{
		Calculated joint transverse-momentum and rapidity spectrum~\(\dNdpTdeta\) of produced charged hadrons (upper solid curve, gray)
		--~averaged for \({\abs{\eta} < \num{0.8}}\) and displayed as function of \(\pT\)~--
		in central \({\sqrtsNN = \qty{5.02}{\TeV}}\) Pb--Pb collisions, compared with \qtyrange{0}{5}{\percent} ALICE data \cite{ALICE2018JHEP2018}.
		Contributions from pions (dashed), kaons (dot-dashed), and (anti)protons (dotted) are shown for the central \(\gluglu\) source, together with the total yield from a single fragmentation source (lower solid curve, blue).
		Vertical bars indicate the statistical plus systematic experimental uncertainties, horizontal bars the bin sizes.
	}
	\label{fig:7}
\end{figure}

At the higher LHC energy of \qty{5.02}{\TeV} in central Pb--Pb collisions, the results of the two-dimensional relativistic diffusion model are
--~apart from the much higher total yield~--
qualitatively similar to the ones at \qty{2.76}{\TeV}, as shown in \cref{fig:6} for the marginal pseudorapidity spectrum~\(\dNdeta\) in comparison with charged-hadron ALICE data \cite{ALICE2017PLB772};
again broken down into pions (dashed curve), kaons (dot-dashed), and protons (dotted) from the central \(\gluglu\) source, plus the total yield from the forward \(\quaglu\) (blue) and backward \(\gluqua\) (red) fragmentation sources.
As in case of the lower LHC energy, the midrapidity dip is overestimated, which we attribute to the same cause.

An average of the joint transverse-momentum and pseudorapidity spectrum for \(\abs{\eta} < \num{0.8}\) is shown in \cref{fig:7}, and compared to ALICE data \cite{ALICE2018JHEP2018} with pion, kaon, and proton contributions for the central source, as well as the small fragmentation contribution arising from the local midrapidity minimum.
As before, the lower frame indicates the normalized residuals
--~measurement minus calculation divided by uncertainty~--
as functions of the plotted coordinate.

Due to the nonmonotonic course of the \qty{5.02}{\TeV} \(\dNdeta\) ALICE measurements around \(\abs{\eta} \approx \num{4}\) and the lack of an experimental joint spectrum with broad phase-space coverage, it proved difficult to obtain a reasonable fit result for the longitudinal diffusivities, as the fit tried to compensate for the unusual features in the data.
Also, in contrast to \qty{2.76}{\TeV}, there are no available measurements of the joint spectrum outside the midrapidity region.
We therefore decided to extrapolate the --~much clearer~-- \qty{2.76}{\TeV} results for \(\parens{\DL \x \incr{t}}_{\gluglu / \quaglu}\) by scaling them with the beam rapidity.
This approach is based on the earlier finding \cite{RoehrscheidWolschin2012PRC86} that the width of the marginal pseudorapidity spectrum is proportional to the logarithm of the respective center-of-mass energy in good approximation.
A similar strategy could be used to predict spectra at future higher incident energies.

\bigbreak

All numerical calculations were carried out using the \texttt{Julia} programming language \cite{BezansonEtAl2017SR59}; solutions of differential equations were performed with the package \texttt{DifferentialEquations.jl} \cite{RackauckasNie2017JORS5} and parameter optimizations with \texttt{Optim.jl} \cite{MogensenRiseth2018JOSS3}.

\section{Conclusions and Outlook}
\label{sec:conclusion}

We have developed a relativistic phase-space diffusion model for the time evolution of the distribution of produced charged hadrons.
It is formulated in three dimensions, which can be parameterized by two momentum-space coordinates due to the cylindrical symmetry of the problem.
For this, we use rapidity-like longitudinal and transverse variables that can be transformed into transverse momentum and pseudorapidity in order to directly compare to data from relativistic heavy-ion collisions.
The theoretical formulation accounts for the transition from the initial state to the observed final state in hadron production.
It is inspired by the phenomenological relativistic diffusion model (RDM) and uses similar key ingredients such as the distinction of two fragmentation sources and a central source that mainly emerges from gluon--gluon collisions, but is based on stochastic particle trajectories constructed from relativistic drift--diffusion processes in phase space.
The initial conditions for the central gluon--gluon source that dominates particle production at LHC energies and for the fragmentation sources, which are shaped by (less productive) interactions of valence quarks with gluons, are obtained from the corresponding parton distribution functions.
Since the approach is based on the principles of stochastic calculus, coordinate transformations of the mathematical objects involved can be easily performed, and the calculations can be done in the most convenient coordinate system, such as transverse and longitudinal rapidity, or transverse momentum and pseudorapidity.

The drift and the diffusion coefficients were determined from mesoscopic considerations, rather than calculating them microscopically.
We have assumed a constant diagonal diffusivity in longitudinal- and transverse-rapidity space, and identify the equilibrium solutions of the resulting Fokker--Planck equation with the particle concentrations in an expanding thermalized gas.
These thermal distributions are shown to be almost reached during the time evolution in the transverse degrees of freedom --~as is usually assumed in thermal models~--, but the system remains far from equilibrium in the longitudinal degrees of freedom.

The two-dimensional fluctuation--dissipation relation that is derived under these premises and fixes the drift coefficient function is discussed in detail and computed numerically.
Since the equilibrium thermal distribution fails to account for high-momentum tails due to the emergence of hard processes, we adapt it to smoothly transition from an exponential to a Pareto distribution, which leads to a modified FDR.

Probability density functions in transverse and longitudinal rapidity are then calculated numerically by solving the FPE for pions, kaons and (anti)protons, transformed to transverse momentum and pseudorapidity space using the full momentum- and mass-dependent Jacobian transformation, and finally combined into marginal and joint spectra to be compared with experimental data from central \qty{2.76}{\TeV} Pb--Pb collisions measured by the ALICE and ATLAS collaborations.
The individual contributions of pions, kaons, and protons from the central gluon--gluon source are calculated and displayed, as well as the contributions from the two fragmentation sources.

Weighted least-square fits were used to determine the model parameters in comparisons with the ATLAS and ALICE data.
Apart from positivity, no constraints were placed on the dimensionless products of transverse and longitudinal diffusivity and the elapsed interaction time between the initial and final state, which determine the degree of thermalization of the respective degrees of freedom.

Convincing agreement with the available experimental joint (\(\dNdpTdeta\)) and marginal (\(\dNdeta\)) spectra for central \qty{2.76}{\TeV} Pb--Pb collisions has been achieved, underlining the validity and usefulness of the model.
Compared with earlier, more phenomenologically oriented one-dimensional RDM-calculations, the yield that is originating from the fragmentation sources is significantly smaller, but not negligible.

The midrapidity dip comes out to be slightly too pronounced in the current model.
Since it originates mostly from the two-dimensional Jacobian transformation from rapidity to pseudorapidity space that is computed without resorting to average momenta or effective masses and the mass gap between pions and nucleons, there is little to no freedom to adjust parameters to improve the results.
The root of the discrepancy is rather found in our assumption of a constant diagonal diffusivity, which may have to be modified using a well-defined momentum dependence.
In addition, an underestimate of the relative contributions of the fragmentation sources caused by unphysical parameterization could amplify the effect.
The same ingredients could also cure the underestimate of the peak in the joint spectrum around \(\pT \approx \qty{300}{\MeV}\) for \(\abs{\eta} < \num{0.8}\).

Because the precision of our numerical solution of the Fokker--Planck equation is not sufficient across multiple orders of magnitude, we concentrate in this work on a comparison with datapoints not smaller than \qty{5}{\percent} of the peak value.
For the data at hand, this effectively confines our results and analyses to the region \(\pT \leq \qty{2}{\GeV}\).
We plan to improve on this in the future by using different solution methods, such as algebraic operator methods.
Since the model is two-dimensional, only central-collision results can be presented -- an extension to noncentral collisions in a fully three-dimensional model is possible, but very involved.
Such an extension would also provide the opportunity to study related phenomena such as elliptic flow.

We have repeated our analysis for the higher incident energy of \qty{5.02}{\TeV} with central Pb--Pb data from ALICE, finding reasonable agreement as in the \qty{2.76}{\TeV} case, with a corresponding overestimate of the Jacobian dip at midrapidity.
However, the lack of a dataset with broad phase-space coverage at this center-of-mass energy made it difficult to determine the free model parameters, so that we had to resort to extrapolation for the longitudinal diffusivities.
In the calculation, less than 80 charged hadrons per unit of pseudorapidity are produced at the beam rapidity -- this is consistent with data taken much earlier in Au--Au collisions at and below \(\sqrtsNN = \qty{200}{\GeV}\) by the PHOBOS collaboration, but it would be necessary to actually obtain LHC forward-physics measurements to confirm our results.
The model can in principle also be used for predictions, such as regarding the forthcoming Pb--Pb run at \(\sqrtsNN = \qty{5.36}{\TeV}\) at the Large Hadron Collider.

\medskip
\textbf{Acknowledgements} \par 
Discussions with Klaus Reygers about the ALICE Pb--Pb data, and Philipp Schulz about corresponding calculations of particle production in asymmetric systems such p--Pb at \qty{5.02}{\TeV} and \qty{8.16}{\TeV} are gratefully acknowledged, as well as constructive remarks of the referees, and discussions in and the hospitality of the Nuclear Theory Group at LBNL Berkeley.

\medskip
\textbf{Data Availability Statement} \par
The experimental data used in this analysis are available on hep-data.
The results of our numerical calculations can be made available upon reasonable request from the corresponding authors.

\medskip
\textbf{Conflict of Interest} \par
The authors declare no conflict of interest.

\appendix
\section{Transformation of the coefficient functions}
\label{app:coefficient-transformation}

Consider the SDE of a generic \(n\)-dimensional stochastic drift--diffusion process~\(\tup{Q} = \tupdelims{\tup{Q}[1],\dots,\tup{Q}[n]}\),
\begin{equation}
	\diff{\func{\tup{Q}[i]}{t}} =
	\func[\big]{\drift[i]{\tup{Q}}}{\func{\tup{Q}}{t}} \x \diff{t}
	+
	\sum_{k=1}^{n} \func[\big]{\diffusion[i k]{\tup{Q}}}{\func{\tup{Q}}{t}} \x \diff{\func{\tup{W}[k]}{t}}
	\quad\text{for}\quad
	i = 1,\dots,n
\end{equation}
with drift coefficient~\(\drift{\tup{Q}}\) and diffusion coefficient~\(\diffusion{\tup{Q}}\), defined via the Stratonovich discretization scheme.
Then, for any coordinate transformation~\(\tup{\varphi}: \mathbb{R}^n \to \mathbb{R}^n\), the object~\(\func{\tup{\varphi}}{\tup{Q}}\) is also an \(n\)-dimensional stochastic drift--diffusion process, and the coefficients of the two processes are connected by
\begin{equation}
	\func[\big]{\drift[k]{\func{\tup{\varphi}}{\tup{Q}}}}{\func{\tup{\varphi}}{\tup{q}}} = \sum_{i=1}^{n} \deriv{\func{\tup{\varphi}[k]}{\tup{q}}}{\tup{q}[i]} \x \func{\drift[i]{\tup{Q}}}{\tup{q}}
	\mathinner{,}\quad
	\func[\big]{\diffusion[k l]{\func{\tup{\varphi}}{\tup{Q}}}}{\func{\tup{\varphi}}{\tup{q}}} = \sum_{i=1}^{n} \deriv{\func{\tup{\varphi}[k]}{\tup{q}}}{\tup{q}[i]} \x \func{\diffusion[i l]{\tup{Q}}}{\tup{q}}
	\quad\text{for}\quad
	k,l = 1,\dots,n
\end{equation}
since the anomalous term in Itô's lemma vanishes.

For example, the transformation from transverse and longitudinal momentum to rapidity with \(\tup{Q} = \tupdelims{\PT,\PL}\) and
\begin{equation}
	\func{\tup{\varphi}}{\pT,\pL} = \tupdelims[\big]{\func{\arsinh}{\pT / m},\func{\artanh}{\pL / E}} = \rap
\end{equation}
yields the partial derivatives
\begin{align}
	\deriv{\func{\tup{\varphi}[1]}{\pT,\pL}}{\pT} &= \frac{1}{\mT}
	\mathinner{,}&
	\deriv{\func{\tup{\varphi}[1]}{\pT,\pL}}{\pL} &= 0
	\mathinner{,}&
	\deriv{\func{\tup{\varphi}[2]}{\pT,\pL}}{\pT} &= - \frac{\pT \x \pL}{\mT^2 \x E}
	\mathinner{,}&
	\deriv{\func{\tup{\varphi}[2]}{\pT,\pL}}{\pL} &= \frac{1}{E}
	\mathinner{.}
\end{align}
As such, choosing a momentum-space diffusion coefficient of the form
\begin{align}
	\diffusion[1 1]{\tupdelims{\PT,\PL}} &= \sqrt{2 \DT} \x \mT
	\mathinner{,}&
	\diffusion[2 1]{\tupdelims{\PT,\PL}} &= \sqrt{2 \DT} \x \frac{\pT \x \pL}{\mT}
	\mathinner{,}&
	\diffusion[1 2]{\tupdelims{\PT,\PL}} &= 0
	\mathinner{,}&
	\diffusion[2 2]{\tupdelims{\PT,\PL}} &= \sqrt{2 \DL} \x E
\end{align}
with \(\DT,\DL \geq 0\)
results in a diagonal rapidity-space diffusion coefficient~\(\diffusion{\Rap} = \sqrt{2} \x \func{\operatorname{diag}}{\DT,\DL}\).

\section{Fast transverse equilibration}
\label{app:fast-transverse-equilibration}

If the transverse diffusivity is significantly larger than the longitudinal one, \(\DT \gg \DL\), the transverse degrees of freedom almost instantly equilibrate for any \emph{given} \(y\) in any timestep of the FPE.
This means that their conditional PDF
\begin{equation}
	\func{\pdf{\func{H}{t} \mid \func{Y}{t}}}{h \mid y} \coloneq \func{\pdf{\func{\Rap}{t}}}{\rap} / \func{\pdf{\func{Y}{t}}}{y}
\end{equation}
should agree in good approximation with the conditional version of \cref{eq:pdf-equ} also for \(t \in \parens{t_{\ini},t_{\fin}}\).
It is then easy to show that it suffices to solve a Fokker--Planck equation for the longitudinal marginal PDF
\begin{equation}
	\func{\pdf{\func{Y}{t}}}{y} \coloneq \int_0^\infty \diff{h} \x \func{\pdf{\func{\Rap}{t}}}{\rap}
\end{equation}
with the Fokker--Planck operator
\begin{equation}
	\func{\fpo{Y}}{y} \coloneq \DL \x \derivop{y} \bracks*{\deriv{\func{\potential{Y}}{y}}{y} + \derivop{y}}
	\mathinner{.}
\end{equation}
The employed generalized potential~\(\potential{Y}\) is computed from the longitudinal marginal PDF of the equilibrium state.
After solving this equation, the joint PDF can be reconstructed through
\begin{equation}
	\func{\pdf{\func{\Rap}{t_{\fin}}}}{\rap} \approx \func{\pdf{\func{Y}{t_{\fin}}}}{y} \x \func{\pdf{H_{\equ} \mid Y_{\equ}}}{h \mid y}
	\mathinner{.}
\end{equation}

\medskip

\IfFileExists{references.bib}{
	\bibliographystyle{MSP}
	\bibliography{references}

\begin{thebibliography}{10}
\providecommand{\url}[1]{\texttt{#1}}
\providecommand{\urlprefix}{URL }

\bibitem{GaleEtAl2013PRL110}
C.~Gale, S.~Jeon, B.~Schenke, P.~Tribedy, R.~Venugopalan,
\newblock \emph{Phys. Rev. Lett.} \textbf{2013}, \emph{110}, 1 012302.

\bibitem{Wolschin2015PRC91}
G.~Wolschin,
\newblock \emph{Phys. Rev. C} \textbf{2015}, \emph{91}, 1 014905.

\bibitem{Mehtar-TaniWolschin2009PRL102}
Y.~Mehtar-Tani, G.~Wolschin,
\newblock \emph{Phys. Rev. Lett.} \textbf{2009}, \emph{102}, 18 182301.

\bibitem{Mehtar-TaniWolschin2009PRC80}
Y.~Mehtar-Tani, G.~Wolschin,
\newblock \emph{Phys. Rev. C} \textbf{2009}, \emph{80}, 5 054905.

\bibitem{HoelckWolschin2020PRR2}
J.~Hoelck, G.~Wolschin,
\newblock \emph{Phys. Rev. Research} \textbf{2020}, \emph{2}, 3 033409.

\bibitem{DebbaschEtAl1997JSP88}
F.~Debbasch, K.~Mallick, J.~P. Rivet,
\newblock \emph{J. Statist. Phys.} \textbf{1997}, \emph{88}, 3/4 945.

\bibitem{DunkelHaenggi2005PRE71}
J.~Dunkel, P.~Hänggi,
\newblock \emph{Phys. Rev. E} \textbf{2005}, \emph{71}, 1 016124.

\bibitem{DunkelHaenggi2005PRE72}
J.~Dunkel, P.~Hänggi,
\newblock \emph{Phys. Rev. E} \textbf{2005}, \emph{72}, 3 036106.

\bibitem{DunkelHaenggi2009PR471}
J.~Dunkel, P.~Hänggi,
\newblock \emph{Phys. Rep.} \textbf{2009}, \emph{471}, 1 1.

\bibitem{Wolschin1999EPJA5}
G.~Wolschin,
\newblock \emph{Eur. Phys. J. A} \textbf{1999}, \emph{5}, 1 85.

\bibitem{Wolschin2016PRC94}
G.~Wolschin,
\newblock \emph{Phys. Rev. C} \textbf{2016}, \emph{94}, 2 024911.

\bibitem{HoelckEtAl2023PLB840}
J.~Hoelck, E.~Hiyama, G.~Wolschin,
\newblock \emph{Phys. Lett. B} \textbf{2023}, \emph{840} 137866.

\bibitem{BiyajimaEtAl2002PTP108}
M.~Biyajima, M.~Ide, T.~Mizoguchi, N.~Suzuki,
\newblock \emph{Prog. Theor. Phys.} \textbf{2002}, \emph{108}, 3 559.

\bibitem{WolschinEtAl2006AP518}
G.~Wolschin, M.~Biyajima, T.~Mizoguchi, N.~Suzuki,
\newblock \emph{Ann. Phys.} \textbf{2006}, \emph{518}, 6 369.

\bibitem{WolschinEtAl2006PLB633}
G.~Wolschin, M.~Biyajima, T.~Mizoguchi, N.~Suzuki,
\newblock \emph{Phys. Lett. B} \textbf{2006}, \emph{633}, 1 38.

\bibitem{Wolschin2013JPG40}
G.~Wolschin,
\newblock \emph{J. Phys. G} \textbf{2013}, \emph{40}, 4 045104.

\bibitem{ALICE2013PLB726}
E.~Abbas, et~al.,
\newblock \emph{Phys. Lett. B} \textbf{2013}, \emph{726}, 4-5 610.

\bibitem{ALICE2017PLB772}
J.~Adam, et~al.,
\newblock \emph{Phys. Lett. B} \textbf{2017}, \emph{772} 567.

\bibitem{ALICE2018JHEP2018}
S.~Acharya, et~al.,
\newblock \emph{J. High Energy Phys.} \textbf{2018}, \emph{2018}, 11.

\bibitem{ATLAS2015JHEP2015}
G.~Aad, et~al.,
\newblock \emph{J. High Energy Phys.} \textbf{2015}, \emph{2015}, 9.

\bibitem{Lopuszanski1953APP12}
J.~Łopuszański,
\newblock \emph{Acta Phys. Polon.} \textbf{1953}, \emph{12} 87.

\bibitem{Dudley1966AM6}
R.~M. Dudley,
\newblock \emph{Ark. Mat.} \textbf{1966}, \emph{6}, 3 241.

\bibitem{Hakim1968JMP9}
R.~Hakim,
\newblock \emph{J. Math. Phys.} \textbf{1968}, \emph{9}, 11 1805.

\bibitem{discretization_scheme}
In this article, all SDEs are defined using the Stratonovich discretization
  scheme, which simplifies coordinate transformations. See
  \cref{app:coefficient-transformation} for more detail.

\bibitem{natural_units}
Throughout this article, we use natural units where \(c = k_{\mathrm{B}} = 1\).

\bibitem{position_independence}
This is not equivalent to the system being homogeneous in position space; this
  is only the case in conjunction with a position-independent initial
  condition.

\bibitem{hyperbolic_transformation}
The latter is of advantage in the computational treatment of the function, such
  as in the numerical solution of its time-evolution equation.

\bibitem{TrainorPrindle2016PRD93}
T.~A. Trainor, D.~J. Prindle,
\newblock \emph{Phys. Rev. D} \textbf{2016}, \emph{93}, 1 014031.

\bibitem{DunkelEtAl2009NP5}
J.~Dunkel, P.~Hänggi, S.~Hilbert,
\newblock \emph{Nat. Phys.} \textbf{2009}, \emph{5}, 10 741.

\bibitem{nonthermal_equilibrium}
The equilibrium state in question need not necessarily be thermal; any
  time-asymptotic stationary state qualifies.

\bibitem{LeeEtAl1990ZPC48}
K.~S. Lee, U.~Heinz, E.~Schnedermann,
\newblock \emph{Z. Phys. C} \textbf{1990}, \emph{48}, 3 525.

\bibitem{Bjorken1983PRD27}
J.~D. Bjorken,
\newblock \emph{Phys. Rev. D} \textbf{1983}, \emph{27}, 1 140.

\bibitem{SchnedermannEtAl1993PRC48}
E.~Schnedermann, J.~Sollfrank, U.~Heinz,
\newblock \emph{Phys. Rev. C} \textbf{1993}, \emph{48}, 5 2462.

\bibitem{ALICE2013PRC88}
B.~Abelev, et~al.,
\newblock \emph{Phys. Rev. C} \textbf{2013}, \emph{88}, 4 044910.

\bibitem{ALICE2020PRC101}
S.~Acharya, et~al.,
\newblock \emph{Phys. Rev. C} \textbf{2020}, \emph{101}, 4 044907.

\bibitem{MichaelVanryckeghem1977JPGNP3}
C.~Michael, L.~Vanryckeghem,
\newblock \emph{J. Phys. G: Nucl. Phys.} \textbf{1977}, \emph{3}, 8 L151.

\bibitem{Hagedorn1983RNC6}
R.~Hagedorn,
\newblock \emph{Riv. Nuovo Cim.} \textbf{1983}, \emph{6}, 10 1.

\bibitem{exponential_sequence}
Not to be confused with the exponential \emph{series}.

\bibitem{ALICE2011EPJC71}
K.~Aamodt, et~al.,
\newblock \emph{Eur. Phys. J. C} \textbf{2011}, \emph{71}, 6.

\bibitem{Gribov1983PR100}
L.~Gribov,
\newblock \emph{Phys. Rep.} \textbf{1983}, \emph{100}, 1-2 1.

\bibitem{MuellerQiu1986NPB268}
A.~H. Mueller, J.~Qiu,
\newblock \emph{Nucl. Phys. B} \textbf{1986}, \emph{268}, 2 427.

\bibitem{BlaizotMueller1987NPB289}
J.~P. Blaizot, A.~H. Mueller,
\newblock \emph{Nucl. Phys. B} \textbf{1987}, \emph{289} 847.

\bibitem{McLerranVenugopalan1994PRD49}
L.~McLerran, R.~Venugopalan,
\newblock \emph{Phys. Rev. D} \textbf{1994}, \emph{49}, 5 2233.

\bibitem{Golec-BiernatWuesthoff1998PRD59}
K.~Golec-Biernat, M.~Wüsthoff,
\newblock \emph{Phys. Rev. D} \textbf{1998}, \emph{59}, 1 014017.

\bibitem{KharzeevEtAl2005NPA747}
D.~Kharzeev, E.~Levin, M.~Nardi,
\newblock \emph{Nucl. Phys. A} \textbf{2005}, \emph{747}, 2-4 609.

\bibitem{MartinEtAl2002PLB531}
A.~D. Martin, R.~G. Roberts, W.~J. Stirling, R.~S. Thorne,
\newblock \emph{Phys. Lett. B} \textbf{2002}, \emph{531}, 3-4 216.

\bibitem{DumitruEtAl2006NPA765}
A.~Dumitru, A.~Hayashigaki, J.~Jalilian-Marian,
\newblock \emph{Nucl. Phys. A} \textbf{2006}, \emph{765}, 3-4 464.

\bibitem{KharzeevEtAl2004PLB599}
D.~Kharzeev, Y.~V. Kovchegov, K.~Tuchin,
\newblock \emph{Phys. Lett. B} \textbf{2004}, \emph{599}, 1-2 23.

\bibitem{BaierEtAl2006NPA764}
R.~Baier, Y.~Mehtar-Tani, D.~Schiff,
\newblock \emph{Nucl. Phys. A} \textbf{2006}, \emph{764} 515.

\bibitem{RoehrscheidWolschin2012PRC86}
D.~M. Röhrscheid, G.~Wolschin,
\newblock \emph{Phys. Rev. C} \textbf{2012}, \emph{86}, 2 024902.

\bibitem{BezansonEtAl2017SR59}
J.~Bezanson, A.~Edelman, S.~Karpinski, V.~B. Shah,
\newblock \emph{{SIAM} Rev.} \textbf{2017}, \emph{59}, 1 65.

\bibitem{RackauckasNie2017JORS5}
C.~Rackauckas, Q.~Nie,
\newblock \emph{J. Open Res. Softw.} \textbf{2017}, \emph{5}, 1 15.

\bibitem{MogensenRiseth2018JOSS3}
P.~K. Mogensen, A.~N. Riseth,
\newblock \emph{J. Open Source Softw.} \textbf{2018}, \emph{3}, 24 615.

\end{thebibliography}
}{

}

\clearpage
\begin{figure}
	\textbf{Table of Contents}\\
	\medskip
	\includegraphics[width=55mm,height=50mm,keepaspectratio]{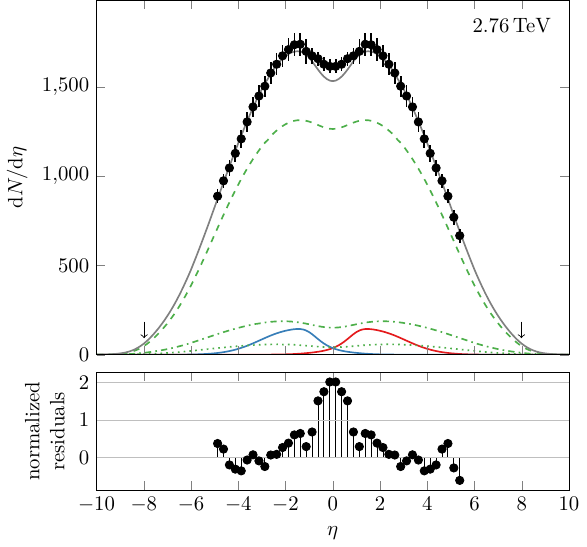}
	\medskip
	\caption*{
		To account for particle production in relativistic heavy-ion collisions at the Large Hadron Collider, a relativistic diffusion model with cylindrical symmetry is derived from nonequilibrium--statistical considerations.
		It propagates an initial state based on quantum chromodynamics in time towards a thermal equilibrium limit.
		Calculated distribution functions in transverse-momentum and pseudorapidity space for pions, kaons, and (anti)protons are compared with recent data.
	}
\end{figure}

\end{document}